\documentclass[12pt]{article}
\usepackage{graphicx}
\usepackage{psfrag}
\usepackage{epstopdf}
\usepackage{latexsym}
\usepackage{slashed}
\usepackage{amsmath, amssymb}

\begin{document}

\begin{titlepage}
\null\vspace{-62pt}

\pagestyle{empty}
\begin{center}

\vspace{0.6truein} {\Large\bf  Confining quantum field theories}

\vspace{0.3in}
{\large Dimitrios Metaxas} \\
\vskip .4in
{\it Department of Physics,\\
National Technical University of Athens,\\
Zografou Campus, 15780 Athens, Greece\\
metaxas@mail.ntua.gr}\\

\vspace{.2in}
\centerline{\bf Abstract}

\baselineskip 18pt
\end{center}

It is widely believed, and axiomatically postulated in mathematical quantum field theory, that the vacuum is a unique vector state.
The recent solution of the quantum Yang-Mills theory of the strong interaction revealed the presence of two vacua and a mixed quantum state.  The second, confining vacuum, is an eigenstate of an auxiliary field, with a non-zero eigenvalue, as opposed to the zero eigenstate of the perturbative vacuum, and provides a new mechanism of scale generation.

I show that this non-trivial vacuum structure implies confinement, in the sense that vacuum expectation values between states separated at large, space-like distances, tend to zero, 
whereas in ordinary quantum theories with a unique vacuum, they are known to satisfy the cluster decomposition principle, and tend to free, asymptotic states, at large separations.

In a confined state, the correlation functions are zero
at spacelike distances larger than the scale of the  theory.
Accordingly, they can be non-zero only along
a timelike worldline (with an associated spacelike width).

The theory is by construction unitary and Lorentz invariant, but the different vacua give a direct sum decomposition.
Implications on determinism and causality, and
generalizations of the confinement mechanism for theories with other symmetries and interactions are discussed.

I argue that confinement, in the generalized sense, is a necessary (certainly not sufficient) condition for proposed theories of a conscious state.
Also, I discuss the relation with the measurement postulate of quantum mechanics
(when the ``observer" is merely a detector).
I argue that confinement, in the strong interaction, is an important mechanism, similar to and possibly along with decoherence, for the measurement process.

\end{titlepage}
\newpage
\pagestyle{plain}
\setcounter{page}{1}
\newpage

\section{Introduction}

In \cite{dkm1}, I used a specific set of Feynman rules, in order to satisfy Gauss's law  and estimate the interaction energy between two static sources in Yang-Mills theory. 
After using an auxiliary operator, a Lagrange multiplier field, $\lambda$, the effective action derived, although at a particular Lorentz  frame, was shown in  \cite{dkm2} to admit two distinct, stable vacua,
the perturbative vacuum, $|\Omega_0>$, with $\lambda =0$, and the confining vacuum, $|\Omega_\mu>$, with $\lambda^2 =\mu^2$, with a dimensionful constant, $\mu$, generated via the Coleman-Weinberg mechanism. 
Physics at each vacuum, as well as in the mixed vacuum state, is Lorentz-invariant. In \cite{dkm3} the role of the confining vacuum as an eigenstate of the gauge-invariant auxiliary operator $\lambda^2 \,|\Omega_\mu> \,=\mu^2 \,|\Omega_\mu>$, was explained,
and contrasted with the ordinary, conventional quantum field theory, with the unique vacuum postulate, which is equivalent to the cluster decomposition property.
A related feature of the confining theory is the fact that $\lambda^2$ is a non-trivial operator, that commutes with all other operators of the theory, but is not identically zero, or a multiple of the identity.

Here, I show that this vacuum structure of the quantum Yang-Mills theory is characteristic of all confining theories, and implies confinement, in the sense that all correlation functions at large spacelike separations tend to zero, instead of satisfying the cluster decomposition property (another  unwarranted postulate of conventional quantum field theory, equivalent to the existence of a unique vacuum).

Also, in the case of Yang-Mills theory, there is a vacuum energy density difference between the two vacua that depends on $\mu^2$, the non-zero eigenvalue of the non-trivial operator, $\lambda^2$. 

This is a new mechanism of scale generation.
Although the combinatorics of the Coleman-Weinberg effective potential are well known, arising from the infrared singularities of the massless, self-interacting, gauge fields, here
there is no spontaneous symmetry breaking. $\lambda$ is an auxiliary field, without a kinetic term, and without any additional degrees of freedom, and $\mu^2$ is an eigenvalue, not a vacuum expectation value from spontaneous symmetry breaking.

The theory is by construction unitary and Lorentz invariant \cite{qft1, qft2, qft3}, but the different vacua give a direct sum decomposition.
I discuss implications on determinism and causality, and possible
generalizations of the confinement mechanism for theories with other symmetries and interactions.

Since in a confined state the correlation functions are zero
at spacelike distances larger than the scale of the  theory,
they can also be shown to be non-zero only along
a timelike worldline (with an associated spacelike width).
This leads to arguments relating confinement with theories of consciousness, but also of quantum measurement.

In Sec.~2, I review the solution of Yang-Mills theory. In Sec.~3, I describe the confining properties of the theory,
and prove the main identity for the correlation functions.
In Sec.~4, I discuss the general structure and properties of confining field theories,
and implications for unitarity and causality. In Sec.~5, I discuss generalizations of the confining mechanism for other theories and symmetries. In Sec~6, I argue for consciousness as a generalized confined state. In Sec.~7, I discuss relations with theories of measurement, and in Sec.~8,
I conclude with some general comments, regarding the general status of quantum field theory and the physical interactions.

\section{The Yang-Mills theory}

The effective action that incorporates Gauss's law in Yang-Mils theory, with an auxiliary Lagrange multiplier field, $\lambda$, was derived as
\begin{equation}
S_{\rm  eff}= \int \frac{1}{2}{E_i^a}E_i^a -\frac{1}{2}{B_i^a}B_i^a +
                        \lambda^a\, {D_i} {E_i^a} + U(\lambda),
\label{eff}
\end{equation}
where 
$E^a_i = F^a_{0i}, B^a_i=-\frac{1}{2}\epsilon^{ijk}F^a_{jk}$,
$F^a_{\mu\nu}=\partial_{\mu}A^a_{\nu}-\partial_{\nu}A^a_{\mu} - g_c f^{abc}A^b_{\mu}A^c_{\nu}$,
for the non-Abelian gauge group with generators  $T^a$, structure constants $f^{abc}$, and coupling $g_c$,
written in terms of the fields
$A_{\mu}= T^a A^a_{\mu}$ , $\lambda=\lambda^a T^a$.

The effective action is gauge invariant under a local gauge transformation 
$\omega(\alpha) = e^{iT^a\alpha^a (x)}$, 
$A_{\mu}\rightarrow\omega A_{\mu} \omega^{-1} + \frac{i}{g_c}\omega\partial_{\mu} \omega^{-1}$,
$\lambda\rightarrow\omega\lambda\omega^{-1}$.

The effective potential $U(\lambda)$ is of the Coleman-Weinberg form, with a minimum at $\lambda^2 = \mu^2$, at a generated scale, $\mu$, but appears inverted in the effective action ($\lambda^2 = \lambda^a \lambda^a$).

Thus, the inverted potential, $-U$, 
has a local minimum at $\lambda=0$, and a global maximum
at $\lambda^2=\mu^2$, but because of Gauss's law and the presence of gauge kinetic terms, the analysis of \cite{dkm2}
demonstrated that they are both stable vacua, 
and in \cite{dkm3} it was shown that they should be interpreted as eigenstates of  the auxiliary,
gauge-invariant operator, $\lambda^2$.

$|\Omega_0>$, with eigenvalue $\lambda = 0$, is the perturbative vacuum, with the usual Coulomb interaction,
and $|\Omega_\mu>$, with $\lambda^2 |\Omega_\mu> = \mu^2 |\Omega_\mu>$, is the confining vacuum.

This is a new mechanism of scale generation, different than spontaneous symmetry breaking. 

Stable soliton solutions of the equations
\begin{equation}
\nabla^2 \lambda^a = \frac{\partial U}{\partial \lambda^a},
\label{sol}
\end{equation}
derived from (\ref{eff}), were also shown to exist, with a spatial extent of the order of $R_{\rm sol} \sim 1/( \alpha_s \mu)$,
where $\alpha_s = g_c^2 /(4\pi)$.

There is an energy density difference between the two vacua:
the vacuum energy density of $\Omega_0$ is zero, and the vacuum energy density of $\Omega_\mu$ is positive (equal to $-U(\mu^2)$). 
The canonical formalism in \cite{dkm2} derives the energy (Hamiltonian) as
\begin{equation}
H =\int d^3x \left( \frac{1}{2} E_i^a E_i^a + \frac{1}{2} B_i^a B_i^a - U\right).
\label{h}
\end{equation}

Both vacua are stable, and there is no well-defined Lagrangian or energy-momentum tensor that connects the two vacua. If there was, it would be possible to perturb one vacuum, using operator configurations from the Lagrangian, and lower the energy of one vacuum, which is not possible.
Transitions between the vacua can only happen in the presence
of finite matter density, or high temperature, via the soliton solutions that connect them.
The strong-CP problem is also easily resolved in the presence of two vacua.
The effective action of (\ref{eff}) is given at a particular Lorentz frame, in order to express the Gauss's law constraint and derive the force between two static sources. The physics at each vacuum, however, and overall, is Lorentz invariant.
At the vacua there is a well-defined energy-momentum tensor, 
$T^{\mu\nu}=T^{\mu\nu}_{\rm YM} -  g^{\mu\nu} U$,
where
$T^{\mu\nu}_{\rm YM}$ is the usual energy-momentum tensor
for perturbative Yang-Mills, and $g^{\mu\nu}$ is the Lorentz metric.
The results can also be seen as a formal justification  of the ``bag model".

The canonical analysis also showed that the auxiliary operator, $\lambda^2$, is a non-trivial operator that commutes with all other operators of the theory, without being identically zero, or a multiple of the identity.
The associated non-trivial, non-unique vacuum structure, and the decomposition into different, distinct vacua, that are eigenstates of this operator, was shown to correspond to fundamental constructions of operator algebras, namely the Gelfand-Naimark-Segal (GNS) theorem.

Also, the breakdown of the cluster decomposition principle, which is equivalent to the uniqueness of the vacuum, was shown to be related to the above structure, as well as the confining properties of the theory.
In the next Section, I show that confinement is characteristic of a theory with a similar, non-trivial, vacuum structure, with different vacua that are eigenstates of a non-trivial operator that commutes with every other operator of the theory.

\section{Vacuum state and confinement}

Quantum field theory deals with local operators, $Q$, defined by smeared out fields, $\Phi$,
\begin{equation}
Q=\int f(x) \, \Phi(x) \, d^4 x,
\label{cpt}
\end{equation}
with smearing functions like $f(x)$ above, that  have support (take non-zero values) at finite regions of four-dimensional Minkowski space-time, for example, of the form $V \times T$, that describe physical operations, measurements of observables, that are performed in the system, at the finite three-dimensional volume, $V$, over a period of  time, $T$.

Pure states of the physical system are described by normalized vectors, $|\Psi>$, of a Hilbert space, and probabilities for physical observations at such pure states are given by inner products, $<\Psi | Q | \Psi>$, whereas
mixed states are described by density matrices.

Translated operators are given by
\begin{equation}
Q(x) = U(x) \, Q \, U^{-1}(x),
\end{equation}
with the unitary $U(x) = e^{i P_\mu \cdot x^\mu}$,
where the operator $P_\mu$ is constructed from the $T_{\mu\nu}$ at each vacuum. 

One can then define operators like \cite{qft2}
\begin{equation}
\tilde{Q} (p) = \int Q(x) e^{-i p\cdot x} d^4 x,
\end{equation}
which changes the energy-momentum of a state vector by $p$,
and 
\begin{equation}
Q(f) =\int Q(x) f(x) d^4 x,
\end{equation}
which shifts the energy-momentum of a state vector by the support, $\Delta$, of 
$\tilde{f} (p) = \int e^{i p \cdot x} f(x) d^4 x$.
Then, if the spectral supports of the state vectors $|\Psi_1>$ and $|\Psi_2>$ are the regions
$E_1$ and $E_2$, respectively (also subsets of the four-dimensional momentum space), we have
\begin{equation}
<\Psi_2 | Q(f) | \Psi_1 > =0\,\,\,\,\,{\rm if}\,\,\,\, (E_1 +\Delta) \bigcap E_2 =\emptyset.
\label{proof1}
\end{equation}

The vacuum of the theory, a vacuum state, $|\Omega>$, is defined by the property that
$<\Omega | Q^*(f) Q(f) | \Omega > =0$, for any operator, $Q(f)$, that lowers the energy (the support of $\tilde{f}(p)$ is not in the positive cone, $p^0 >0$, $p^2 >0$).

Once the vacuum state is defined and established, any other state of the theory can be built from it,
in the sense that it can be approximated as $Q |\Omega>$, with arbitrarily large precision, where $Q$ is an operator
of the theory. Namely, the vacuum is a cyclic state that defines the theory. This is either postulated or built in the GNS construction.

It is usually assumed that the vacuum is a unique (up to a phase) vector state.
However, the recent solution of Yang-Mills theory, demonstrated the existence of the two aforementioned vacua,
that are eigenstates of the operator $\lambda^2$, which commutes with all other operators of the theory.

I will show here that this structure implies confinement in a rather general way, such that, for any two local operators,
and any configuration of spacelike separated points at equal time, $x_1 =(t, \vec{x}_1)$, $x_2=(t, \vec{x}_2)$,
\begin{equation}
<\Omega_\mu | Q_1 (x_1) Q_2 (x_2) | \Omega_\mu>\,\,\, \rightarrow \,\, 0,
\label{conf1}
\end{equation}
at large spacelike separation $R=|\vec{x_1}-\vec{x_2}|$.

This is opposed to the situation of ordinary, conventional, quantum field theory, with a unique vacuum vector, $|\Omega>$, where
the limit in the right-hand side of (\ref{conf1}) is a product of free states
\begin{equation}
<\Omega | Q_1 (x_1) Q_2 (x_2) | \Omega>\,\,\, \rightarrow \,\, 
<\Omega | Q_1 (x_1) | \Omega><\Omega | Q_2 (x_2) | \Omega>.
\label{cluster}
\end{equation}

Obviously, (\ref{cluster}) is  the cluster decomposition property, which expresses the fact
that two interacting states of a system, when separated at large distances become non-interacting.
This is true for theories like quantum electrodynamics, but cannot possibly be expected to hold for confining theories, like the strong interaction.

In fact, I will prove a stronger statement, namely that
\begin{equation}
<\Omega_\mu | A \, Q_1(x_1) Q_2(x_2) \, B |\Omega_\mu>\,\,\, \rightarrow \,\, 0,
\label{conf2}
\end{equation}
at large spacelike separation $R=|\vec{x_1}-\vec{x_2}|$, for any operators $A, B$, of the theory, and then 
generalize it for other correlation functions of the theory.

First, one notes that any operator, $Q$, can be decomposed in three parts, as
$Q= Q^- + Q^+ + Q^0$, 
with the spectral supports of $Q^+, Q^-$ on the forward and backward light cones respectively,
 and with the spectral support of $Q^0$
being a bounded subset of the light cones of the energy-momentum space
with $|p^0| \leq c$, for any arbitrary positive constant, $c$.

Indeed, after defining $\tilde{Q} (p) = \int f(x-y) \Phi(x) e^{-i p\cdot y} d^4 y \,d^4 x$, for any $Q=\int f(x) \Phi(x) d^4 x$,
and using a partition of unity, $F_+ + F_- +F_0 =1$, with 
\begin{eqnarray}
&& F_+(p) =0 \,\, {\rm for} \,\, p^0 < b, \nonumber \\
&& F_- (p) = 0 \,\, {\rm for} \,\, p^0 > -b \nonumber \\
&& F_0 (p) = 0 \,\, {\rm for} \,\, |p^0| >c \nonumber \\
&& 0 < b < c,
\end{eqnarray}
$Q=\int (F_+(p) +F_- (p) + F_0 (p)) \tilde{Q} (p) d^4 p $ satisfies the criteria for the required splitting.

Then, with this decomposition in the operators,
the expression in (\ref{conf2}) becomes
\begin{equation}
<\Omega_\mu | (A^- + A^0) \, (Q^-_1 +Q^0_1+Q_1^+)  (Q^-_2  +Q^0_2 + Q^+_2) \, (B^+  +B^0) |\Omega_\mu>.
\end{equation}
and, for large $R$, one can commute $Q_1^+ $ to the right and $Q_2^-$ to the left in this expression.
The resulting vectors have spectral support $p^0 > -c$ (the vector at the right side)
and $p^0 < c$ (the vector at the left).

Now, $|\Omega_\mu>$ is an eigenvector of the operator $\lambda^2$, with a non-zero eigenvalue, and  commutes with any operator. So, for distances $R$ that are larger than the soliton radius, $R_{\rm sol}$,
and taking the constant, $c$, of the analysis to be smaller than the soliton energy, $E_{\rm sol}$,
we see that we can place a soliton operator, $Q_{\rm sol}$,  between $0$ and $R$, between the two vectors in the
expression of (\ref{conf2}), and shift the spectra of the vectors so that they do not overlap.
The soliton operator, being a solution of (\ref{sol}), is composed of the $\lambda$ operator and commutes with other operators.
Then, using (\ref{proof1}), we get the zero result for the expression in (\ref{conf2}).

There is a straightforward generalization of this result to a statement for general correlation functions of the theory,
namely that
\begin{equation}
<\Omega_\mu| T(Q_1(x_1) Q_2(x_2) ... Q_N(x_N))|\Omega_\mu> =0,
\label{basic}
\end{equation}
(for any $x_i, N$) unless the $x_i$ lie in a timelike worldline (with a spacelike width of the order of the scale of the theory).

Implications of this statement will be discussed in the next Sections. It should be noted that its proof used the extension of the Hilbert space of the theory from ${\cal H}_0 \oplus {\cal H}_\mu$, build on the asymptotic and confining vacua
 $\Omega_0$ and $ \Omega_\mu$, to ${\cal H}_0 \oplus {\cal H}_\mu \oplus {\cal H}_{\rm sol}$,
that includes the soliton Hilbert space.
It is possible to derive the same statement solely in ${\cal H}_0 \oplus {\cal H}_\mu$,
instead of placing a soliton operator between spacelike separated points, we can consider an energy increasing tube of the confining vacuum,
making essentially a statement for the bag model, which expresses the fact that the strong interactions are confined at the scale of the theory for energy reasons.

\section{Confining and asymptotic quantum field theories}

Conventional quantum field theories \cite{qft1, qft2, qft3} (typical examples being quantum electrodynamics with a coupling $e^2$, or a single scalar field, $\Phi$, with quartic coupling $g\Phi^4$ and without symmetry breaking) consist of a unique vacuum state, $|\Omega>$,
and operators, $Q$, localized in regions of spacetime.
Physical states are represented by vectors, $|\Psi>$, of the Hilbert space, $H$, of the theory, and mixed states by density matrices.
Every vector state, $|\Psi>$, of the physical system of such theories can be built (approximated) by operators acting on the vacuum (the vacuum is cyclic). That is, one can find operators, $Q$, such that $|\Psi>\, \approx Q |\Omega>$, at any level of approximation, for any $|\Psi>$  (the set of all such $Q |\Omega>$ is dense in $H$).

It is easy to see then, that in such theories, any operator $\lambda$, that commutes with all other operators of the theory,
is trivial (either zero or a multiple of the identity).

The structure of the vacuum, even in these theories, however, is far from trivial (the main reason being the cyclicity property).
Physically, the vacuum state contains all the vacuum-to-vacuum diagrams and transitions, all closed loop Feynman diagrams, so it is different for different values of the coupling, and accordingly, the Hilbert space of the theory depends on the coupling and is different from the ``free" theory vacuum  with zero coupling (Haag's theorem).
Free, asymptotic states, however, can be built, and an $S$-matrix can be defined, essentially based on the cluster decomposition property, which is equivalent to the uniqueness of the vacuum.

The relation of these two properties, which is, actually, a mathematical, logical equivalence (uniqueness of the vacuum $\Leftrightarrow$ cluster decomposition property) was observed as early as in \cite{cdp}, it is noted in \cite{qft1}, and is also mentioned in \cite{qft2, qft3} in connection with results from operator algebras, such as the GNS construction.
All these, as well as subsequent works, however, and eventually quantum field theory textbooks, adopt them as a postulate (a mathematical axiom) in order to build and justify the $S$-matrix construction, which was used to describe the experimental results investigating particle physics.
In fact, the early related theorems \cite{cdp} were derived even before the proposal of the quark model, which was  later included in the Standard Model, and its eventual establishment. 

The strong interaction, however, with a linearly rising interaction, and confinement, cannot possibly be expected to satisfy the cluster decomposition property, and thus lead to a similar construction. There are no asymptotic states, and there is no such thing as an $S$-matrix involving elementary gluons and quarks at distances larger than the scale of the theory.
An interesting and useful discussion of relative scales and confining theories at various limits appears in \cite{georgi}, although it was in a different context, and should be interpreted now with the condition of the mixed vacuum state.

In the strong interaction described by Yang-Mills theory,
the vacuum is a mixed state, with the asymptotic and confining vacua $\Omega_0$ and $ \Omega_\mu$,
and the Hilbert space is a direct sum  ${\cal H}_0 \oplus {\cal H}_\mu$,
so that the Hamiltonian of the theory is also decomposed as
\begin{equation}
\begin{pmatrix}
H_0 &0\\0 &H_\mu
\end{pmatrix}
\label{h1}
\end{equation}
The existence and stability of the solitons connecting the two vacua
enlarges the Hilbert space to  ${\cal H}_0 \oplus {\cal H}_\mu \oplus {\cal H}_{\rm sol}$,
and the interactions can be schematically described by the Hamiltonian
\begin{equation}
\begin{pmatrix}
H_0 &0 &H_{\rm s 0}\\0 &H_\mu &H_{\rm s 1}\\ H_{\rm s 0} &H_{\rm s 1} & H_{\rm sol}
\end{pmatrix}
\label{h2}
\end{equation}
From the previous analysis, only $H_0$ and $H_\mu$ have been completely determined, as in (\ref{h})
in the work of \cite{dkm2}, with the appropriate commutation relations
and Feynman rules. The interactions involving solitons, $H_{\rm sol}$, and the off-diagonal terms, can be calculated in principle, but the analytical treatment is 
limited to large soliton separations.

It should be stressed that the interactions are unitary and Lorentz invariant. Because of confinement, however, as 
expressed in general in (\ref{basic}),
and as we know from everyday practice, the manifestation and verification of causality and unitarity are performed 
in mixed and composite states of baryons, protons, etc., instead of elementary quark and gluon states.

We have a unitary evolution, as described schematically in (\ref{h1}) or (\ref{h2}),  but the structure of the vacuum, along with the confinement properties, already pose important conceptual questions regarding initial conditions, Cauchy surfaces, and determinism. The main problem is that, practically and experimentally, we can claim to be able to fix initial conditions, at the fundamental level, in the perturbative, asymptotic states, that lie in ${\cal H}_0$,  but only in composite confined states when it comes to ${\cal H}_\mu$, as explained above. 

Obviously, since the entire evolution is unitary and Lorentz invariant (in fact, these are the basic requirements built in the algebraic construction described in the next Section)
it remains so if the initial and final states are confined (protons, baryons, hadrons)
but this is not checked at the level of quarks and gluons, only at the level of the composite initial and final states.
Intermediate states and calculations also obey unitarity and causality, by (\ref{h1}) or (\ref{h2}), but the experimental setups, preparations and verifications are on composite states.

\section{Generalizations}

In order to gain more insight, trust the process, and venture in other demonstrations and generalizations of confining phenomena, it is useful to recall the axiomatic algebraic quantum field theory that uses operator algebras in the foundations of its construction.

In fact, the starting point is even more abstract than operator theory, because one starts with a general ${\cal C}^*$- algebra, ${\cal A}$ over the complex numbers, consisting of elements, $A$, admitting conjugates, $A^*$, satisfying properties such as $(A_1 A_2)^* =A_2^* A_1^*$,
with a norm obeying $|| A^* A || = ||A||^2 $, complete with respect to the norm.

A state of ${\cal A}$ is a linear functional, $\phi$, from ${\cal A}$ to the complex numbers, which is positive,
$\phi(A^* A) \ge 0$, and normalized, $\phi(I) =1$, where $I$ is the unity element of the algebra.

Once a state is given, there is a construction (GNS) that associates and builds a Hilbert space, ${\cal H}$, related to the state and the algebra, and provides a representation of the ${\cal C}^*$-algebra as an algebra of continuous operators on ${\cal H}$, and makes contact with the foundations of quantum mechanics and its observables as self-adjoint elements of the algebra and the related operator space. In the case of quantum field theory, the abstract elements of the algebra can be related to operators
of the form (\ref{cpt}).

Symmetries are expressed by elements, $g$, of a Lie group, $G$, and the action of a symmetry on the elements of the algebra transforms every operator $Q$ to $U_g Q U_g^*$, with a unitary representation, $U_g$, of the group on the Hilbert space.

Quantum field theory considers the observables measured in the spacetime domain $D = V \times T$, of the finite three-dimensional volume, $V$, of a laboratory, over a finite period of  time, $T$.
These observables, ${\cal A}(D)$, over all compact subsets, $D$, of Minkowski spacetime, generate an abstract algebra satisfying
the basic postulates of locality (on spacelike separated distinct subregions, they commute)
and Poincare covariance.

The Poincare group, ${\cal P}$, is the semidirect product, 
\begin{equation}
{\cal P} = {\cal T} \ltimes {\cal L},
\label{general}
\end{equation}
of the spacetime translations group, ${\cal T}$,
and the Lorentz group, ${\cal L}$, that preserves the spacetime metric.

The translation group has a unitary representation $T(a) = e^{i P \cdot a}$, for any vector, $a^{\mu}$, of the Minkowski spacetime,
and $P^{\mu}$ is the energy-momentum vector. Time evolution is done with the unitary Hamiltonian operator, $P^0$.

Operators that decrease the energy are defined as in Section~3, and the vacuum state, $\phi$, is defined, so that it satisfies $\phi(Q^* Q) =0$, for any operator, $Q$, which decreases the energy in some coordinate system.

Unitarity and Poincare invariance are built in the construction of the quantum field theory, which is generated with the GNS prescription from the vacuum state.
In the case of the Yang-Mills theory, the vacuum state is not a pure vector state, it is mixed, and the resulting Hilbert space is a direct sum of Hilbert spaces built on the perturbative and the confining vacuum, $|\Omega_0>$ and $|\Omega_\mu>$, respectively.

In the confining vacuum we have (\ref{basic})
\begin{equation}
<\Omega_\mu| T(Q_1(x_1) Q_2(x_2) ... Q_N(x_N))|\Omega_\mu> =0,
\end{equation}
for any operators, at any spacetime points, $x_i$, unless they lie in a timelike worldline (with a spacelike width of the order of the scale of the theory).
Spacelike confinement leads to a timelike evolution along, what is sometimes called, a worldline of an observer.

The notion of an observer is not well-defined in physics, it is sometimes used to denote a measuring device (a detector making measurements of physical observables) and sometimes to denote a conscious entity registering such events.
I will argue, based on (\ref{basic}), that these two very different concepts use the property of confinement as a necessary (certainly not sufficient) condition.

\section{Consciousness as a confined state}

There are several competing theories of consciousness, and they sometimes have some common postulates. Instead of comparing and analysing various different approaches, I will first make some comments using some of the postulates of the integrated information theory of consciousness (IIT) \cite{tononi} and my main point is to argue for confinement as a necessary condition,
its relevance, and possibility of realization in a generalised context, not to present a new, or justify an existing theory of consciousness.

According to  some of the postulates of IIT, consciousness has the properties of integration and exclusion.
Namely, it is irreducible to separate elements, it has borders, and is unified.
The integrated information of the system can be thought of as information that is not stored in spacelike separated partitions, but ``integrated" in the system as a whole.
It is also interesting to note other arguments of theories of consciousness that challenge views of memory as stored in spacelike places in the brain \cite{arcaya}, and propose ``temporal" structure and organization.

The above characteristics of some of the postulates of IIT and other theories of consciousness are naturally embodied in a confined state, with the basic property (\ref{basic})
\begin{equation}
<\Omega_\mu| T(Q_1(x_1) Q_2(x_2) ... Q_N(x_N))|\Omega_\mu> =0,
\end{equation}
unless the $x_i$ lie in a timelike worldline (with an appropriate spacelike width).

Neither IIT, nor any other theory of consciousness, can easily accomodate the cluster decomposition property
that holds on asymptotic states built on a unique vacuum.

Interestingly, there have also been some arguments for theories of consciousness using quantum mechanical treatments \cite{nanopoulos}.
These arguments use quantum mechanics, both conceptually and physiologically, in order to 
quantify and qualify information, but they do not consider confinement in the sense presented here.

I would like to extend the quantum arguments, using the more general case of the algebraic method presented in Section~5, with the dynamics of operator algebras and symmetries, such as (\ref{general}).

Namely, operator algebras express interactions and dynamics that were considered to  describe relativistic quantum mechanics, but this does not have to be their only realization.

One may entertain the possibility of more general dynamics (noncommutative and highly nontrivial, of course) and other associated symmetries (but with a ``temporal" and ``spatial" component) which are realized and admit a mixed vacuum state that enjoys spatial confinement.

The dynamics of the operator algebra, the symmetries
${\cal P} = {\cal T} \ltimes {\cal L}$, their representation $T(a)=e^{i P\cdot a}$, and temporal evolution
governed by $P^0$, do not have to correspond to a quantum interaction,
or the Lorentz and Poincare group.
They can be related to different interactions and symmetries (quantum, classical, neural, biological)
as long as they have a generalised operator algebra realization,
a mixed, confining vacuum state,
and distinct temporal and spatial components.

Then, the state of spatial confinement, as expressed with relations similar to (\ref{basic}),
after a period which would be the equivalent of the biological evolution, can be considered to lead to a state of temporal consciousness of the kind experienced.

Several conceptual, and also some metaphysical questions can be better understood with the paradigm of ``asymptotic" and ``confined" states suggested here.
For example, one can claim that ``robot" or ``zombie" states (defined as states that are assembled from asymptotic data, and programmed to react in a multitude of external stimuli with the same response as a conscious state) are not conscious, since they are, by default, asymptotic, not confined.
 The initial value problems that were also mentioned at the end of Section~4 can also be related to conceptual and metaphysical questions regarding ``free will", since
conscious entities use sensors for all communications, and these respond to asymptotic data.
The confinement paradigm shows that practical setups and interactions involving only asymptotic states (photons and electrons, for example)
cannot fully describe a confined state (neither its initial condition nor its evolution).
Providing initial data on ${\cal H}_0$, does not fully determine the evolution on ${\cal H}_0 \oplus {\cal H}_\mu$.

\section{The measurement problem}

Most interpretations and practitioners of quantum physics agree that the concept of quantum measurement is problematic, not fully explained, and perhaps not even described within the framework of the theory itself, possibly associated with a non-unitary evolution that cannot be accomodated by its premises.

Important steps towards resolving the problem of measurement and wavefunction collapse have been made with 
the concept of decoherence \cite{zurek}, which considers the evolution of the quantum system to be measured, with states $|\psi_n>$,
together with the measuring device, with pointer states $|\chi_n>$, and a reservoir with states $|\phi_n>$.

After a period of interaction, the state 
\begin{equation}
\sum_n c_n |\psi_n> |\chi_n> |\phi_n>,
\end{equation}
evolves to the density matrix
\begin{equation}
\sum_{n, n'} c_n c^*_{n'} <\phi_{n'}(t) | \phi_n (t)> |\psi_n, \chi_n><\psi_{n'}, \chi_{n'}|,
\end{equation}
after a partial trace over the reservoir states.
It is claimed that the large size of the reservoir causes the non-diagonal matrix elements
$<\phi_{n'}(t) | \phi_n (t)>$ to decay to zero, and reduces the 
final state to a diagonal density matrix that corresponds to a statistical mixture.

The observation made here, based on (\ref{basic}), is that the desired relation, $<\phi_{n'}(t) | \phi_n (t)> =0$,
for $n \neq n'$, can be related to confinement. Although usually present, there is no need for a separate and ``large" reservoir, these are the confined 
states of the different parts of the detector. 

Here the different parts of the detector are considered confined by the usual strong interaction.
There is no need for the generalised concepts of confinement described in the previous Section,
and certainly no need for the concept of consciousness in order to make a measurement.
Conscious entities, on the other hand, also have physical parts, consisting of ordinary matter and responding to external stimuli,
and can certainly make measurements, which will also look like a statistical mixture to them.

One may look at this argument as a further justification of decoherence, or as a process that takes place along with decoherence, to the same effect. It can be related to the emergence of classical behavior and the collapse of the wavefunction, in a similar manner. One can also consider ``metatheorems" such as the statement that we cannot build a detector out of just asymptotic matter, say, just electrons and photons; we also need the baryons that bind them.

\section {Conclusion}

As in the original work \cite{dkm1},
the comments and results presented in \cite{dkm2, dkm3} and here are somewhat disparate.
The reason is that confinement is related to a generalization of the
basic axioms of quantum field theory, and this can be  associated with several distinct phenomena.

It is also hoped that the problems of quantum gravity can be related to these investigations,
that the constraints of the spin-$2$ gravitons can be treated in a similar manner as the spin-$1$,
non-Abelian gauge bosons, and that strong gravitational and cosmological effects, involving black hole
or de~Sitter horizons, can be clarified.

The postulate of the uniqueness of the vacuum is made in the original formulation
of axiomatic quantum field theory, in terms of correlation functions \cite{qft1}.
In the algebraic formulation \cite{qft2, qft3}, briefly discussed here in Section~5,
only the physical postulates of causality and Lorentz invariance are used in the foundations.
The vacuum can be a mixed state, and the postulate of uniqueness is made in order to 
have the additional, equivalent property of cluster decomposition, which states that, when we pull a physical system
apart, we reduce it to various, weakly interacting, spacelike separated components.

This is not what happens in the strong interaction. It is noted, however, that the strong interaction is part of the Standard Model,
and many precise estimates have to be made, that involve the confining vacuum, combining contributions from theoretical, phenomenological, experimental, and lattice physics.
The results of the work presented here can be thought of as justifications of older phenomenological models, such as the ``bag model", with the introduction of only a few parameters, the confining vacuum energy density and the mixing between the two vacua.
The breakdown of conventional quantum field theory at the axiomatic level, starting from ``axiom zero",
raises many problems and questions towards the treatment of the strong interaction,
from the theoretical and axiomatic level of Yang-Mills theory, to the calculation of its
effects at the various scales involved.

\section*{\centering Note}
 There are no additional data and no conflicts of interest regarding this work.

\end{document}